\documentstyle[preprint,aps,pre]{revtex}
\begin{document}
\tighten
\title{Tracer diffusion in granular shear flows}
\author{Vicente Garz\'{o}}
\address{Departamento de F\'{\i}sica, Universidad de Extremadura, E-06071 \\
Badajoz, Spain}

\date{\today}
\maketitle

\begin{abstract}
Tracer diffusion in a granular gas in simple shear flow is analyzed. The 
analysis is made from a perturbation solution of the Boltzmann kinetic equation 
through first order in the gradient of the mole fraction of tracer 
particles. The reference state (zeroth-order approximation) corresponds to a 
Sonine solution of the Boltzmann equation, which holds for arbitrary 
values of the restitution coefficients. Due to the anisotropy induced in the 
system by the shear flow, the mass flux defines a diffusion tensor 
$D_{ij}$ instead of a scalar diffusion coefficient. The elements of this 
tensor are given in terms of the restitution coefficients and mass and 
size ratios. The dependence
of the diffusion tensor on the parameters of the problem is 
illustrated in the three-dimensional case. The results show that the 
influence of dissipation on the elements $D_{ij}$ is in general quite 
important, even for moderate values of the restitution coefficients. 
In the case of self-diffusion (mechanically equivalent particles), the 
trends observed in recent molecular dynamics simulations are similar to 
those obtained here from the Boltzmann kinetic theory. 
\end{abstract}

\draft

\bigskip \narrowtext

\newpage

\section{Introduction}
\label{sec1}

Granular systems under rapid flow conditions can be modeled as a fluid of inelastic 
hard spheres. In the simplest model the grains are taken to be smooth so 
that the inelasticity 
is characterized through a constant coefficient of normal restitution.
The essential difference with respect to molecular fluids is the absence of energy 
conservation yielding modifications of the usual hydrodynamic equations. Due to the 
kinetic energy dissipation in collisions, energy must be externally injected 
to the granular gas in order to achieve a stationary state. 
In some experimental situations, the granular system is 
driven into flow by the presence of a shear field. In this case, a steady 
state is possible when the amount of energy supplied by shearing work is 
balanced by that lost due to the inelastic cooling. 
The study of the rheological properties of this steady 
shear flow state has received a great attention in the past years, especially in the 
case of one-component systems\cite{C90}.
However, much less is known in the more complicated case
of multicomponent mixtures of grains.

An interesting problem is the analysis of diffusion in granular shear flows. 
The understanding of mass transport in granular systems is of practical interest 
since, for instance, powders must frequently be mixed togheter before any 
sort of processing can begin. The self-diffusion phenomenon in granular 
flows has been studied earlier. Experimental studies include 
both systems with macroscopic flows \cite{NHT95,MD97} and vertically 
vibrated systems\cite{ZS91}. Complementary computer simulation studies have 
been also carried out\cite{SD93}, with special emphasis on the influence 
of the solid volume fraction on the diffusive motion of the grains. 
In general, all these previous studies were limited to observing the 
diffusion in only one direction, usually the direction parallel to the
velocity gradient. However, due to the anisotropy induced in the system by 
the presence of shear flow, a diffusion tensor is required to describe the 
diffusion process instead of a single diffusion coefficient. To the best of 
my knowledge, the only attempt to measure the elements of this 
self-diffusion tensor has been made by Campbell\cite{C97}. He measured these 
elements via molecular dynamics simulations by using both particle tracking 
and through velocity correlations. Both methods were found to agree with 
reasonable accuracy. Very recently\cite{HY02}, the self-diffusion 
coefficients have been experimentally measured in a granular system under 
Couette flow by employing image technology.

In the context of kinetic theory, studies on granular flows in mixtures are 
scarcer. Most of them \cite{JM89} are based on a Navier-Stokes 
description and, therefore, they are restricted to small velocity 
gradients, which for the steady simple shear flow is equivalent to the 
low-dissipation limit. For this reason, the diffusion is only characterized 
by a single coefficient which is not affected by the presence of the shear 
field. In addition, although these studies permit in 
principle different temperatures for the two species, they assume equal 
partial granular temperatures $T_i$ in the quasielastic limit. 
Nevertheless, given the intrinsic connection between the shear rate and 
dissipation in this problem, energy nonequipartition is expected as the 
restitution coefficient decreases. As a matter of fact, some recent results 
obtained in molecular dynamics simulations of granular sheared 
mixtures\cite{CH02} as well as in real experiments of vibrated mixtures in 
three \cite{WP02} and two dimensions \cite{FM02} clearly show the 
breakdown of energy equipartition. This implies that the temperatures $T_i$ 
are different from the mixture temperature $T$. The consequences of this 
effect on the transport properties are in general significant, as has been 
recently found in the freely cooling case\cite{GD99,GD02}. In conclusion, a 
consistent theory describing diffusion in granular shear flows must take 
into account both the tensorial character of the mass transport as well as   
the possibility of temperature differences.

The aim of this paper is to get the diffusion tensor 
in a binary granular mixture under simple shear flow in the framework of the 
Boltzmann equation.  Due to the complexity of the general problem, here we 
consider the special case in which one of the components (say for instance, 
the species 1) is present in tracer concentration. The tracer problem is 
more amenable to analytically treatment. First, the tracer particles are 
directly enslaved to the granular gas and, second there are fewer 
parameters\cite{DG01,SD01,DBL02}. Therefore, in this situation 
one can assume that the velocity distribution function $f_2$ of the excess
component (granular gas) obeys a (closed) nonlinear Boltzmann equation while 
the velocity distribution function $f_1$ of the tracer particles 
(impurities) satisfies a Boltzmann-Lorentz equation. The starting point 
of our study is a recent solution of the set of Boltzmann equations for a 
binary mixture of inelastic hard spheres  under shear flow\cite{MG02}. 
The corresponding Boltzmann-Lorentz equation for the impurities is solved by 
means of a perturbative scheme in powers of the gradient of the mole 
fraction of tracer particles. The main feature of this expansion is 
that the reference state around which we perturb is not restricted to 
small values of the shear rate, which for the steady shear flow problem is 
equivalent to arbitrary degree of dissipation. In the first order of the 
expansion, the tracer diffusion tensor is identified from the mass flux. 
Explicit expressions for the nonzero elements of this tensor are obtained by 
using a first Sonine polynomial approximation. These elements are given in 
terms of the restitution coefficients and the parametes of the mixture 
(masses and sizes).

The plan of the paper is as follows. In Sec.\ \ref{sec2} we introduce the 
set of coupled Boltzmann equations describing the mixture and 
state the problem we are interested in. The state of
the mixture in the absence of diffusion is analyzed and in particular,
the nonzero elements
of the pressure tensor ${\sf P}_2$  of the granular gas are obtained in 
the leading Sonine approximation. Comparison of these results with 
previous theories is also presented. The Section ends studying the state of
tracer particles with special emphasis in the evaluation of
the specific dependence of the temperature
ratio $T_1/T_2$ on restitution coefficients, mass ratio, and size ratio.
Section \ref{sec3} deals with the perturbation scheme used to solve
the Boltzmann-Lorentz equation of the tracer particles when the diffusion
takes place in the system. This Section contains the main results of
the paper since we determine the tracer diffusion tensor in the first order of the expansion of the 
concentration gradient. The dependence of the nonzero elements of this
tensor on the different parameters of the problem is 
illustrated in the three dimensional case, showing a good qualitative 
agreement with Campbell's simulations \cite{C97}. 
Finally, in Sec.\ \ref{sec4} we close the paper with some concluding 
remarks.

\section{Description of the problem: Granular mixture in simple shear flow}
\label{sec2}

We consider a granular binary mixture composed by smooth inelastic disks 
($d=2$) or spheres ($d=3$) of masses $m_{1}$ and $m_{2}$ and diameters 
$\sigma_{1}$ and $\sigma_{2}$. Collisions between particles are inelastic 
and characterized by three constant restitution coefficients $\alpha_{11}$, 
$\alpha_{22}$, and $\alpha_{12}=\alpha_{21}$, where $\alpha_{ij}\leq 1$ 
refers to the restitution coefficient for collisions between particles of 
species $i$ and $j$. In the low-density regime, the one-particle velocity 
distribution functions $f_i({\bf r},{\bf v}_1;t)$ ($i=1,2$) obey the set of 
nonlinear Boltzmann kinetic equations: 
 \begin{equation}
\label{2.1}
\left(\frac{\partial}{\partial t}+{\bf v}_1\cdot \nabla \right)f_{i}
({\bf r},{\bf v}_1;t)
=\sum_{j}J_{ij}\left[{\bf v}_{1}|f_{i}(t),f_{j}(t)\right] 
\;,
\end{equation}
where the Boltzmann collision operator 
$J_{ij}\left[{\bf v}_{1}|f_{i},f_{j}\right]$ describing the scattering of 
pairs of particles is 
\begin{eqnarray}
J_{ij}\left[{\bf v}_{1}|f_{i},f_{j}\right]  &=&
\sigma _{ij}^{d-1}\int d{\bf v}_{2}\int d\widehat{\bbox {\sigma }}\,\Theta 
(\widehat{\bbox {\sigma}}\cdot {\bf g}_{12})(\widehat{\bbox {\sigma }}\cdot 
{\bf g}_{12})  \nonumber \\
& &\times \left[ \alpha_{ij}^{-2}f_{i}({\bf r},{\bf v}_{1}',t)f_{j}(
{\bf r},{\bf v}_{2}',t)-f_{i}({\bf r},{\bf v}_{1},t)f_{j}(
{\bf r},{\bf v}_{2},t)\right] 
\;. 
 \label{2.2}
\end{eqnarray}
Here, $d$ is the dimensionality of the system, $\sigma_{ij}=\left( 
\sigma_{i}+\sigma_{j}\right)/2$, $\widehat{\bbox {\sigma}}$ is a unit vector 
along their line of centers, $\Theta$ is
the Heaviside step function and ${\bf g}_{12}={\bf v}_{1}-{\bf v}_{2}$. In 
addition, the primes on the velocities denote the initial values $\{{\bf 
v}_{1}^{\prime},
{\bf v}_{2}^{\prime}\}$ that lead to $\{{\bf v}_{1},{\bf v}_{2}\}$
following a binary collision: 
\begin{equation}
\label{2.3}
{\bf v}_{1}^{\prime }={\bf v}_{1}-\mu_{ji}\left( 1+\alpha_{ij}    
^{-1}\right)(\widehat{\bbox {\sigma}}\cdot {\bf g}_{12})\widehat{\bbox 
{\sigma}},
\quad {\bf v}_{2}^{\prime}={\bf v}_{2}+\mu_{ij}\left( 
1+\alpha_{ij}^{-1}\right) (\widehat{\bbox {\sigma}}\cdot {\bf 
g}_{12})\widehat{\bbox{\sigma}}\;,  
\end{equation}
where $\mu_{ij}=m_{i}/\left(m_{i}+m_{j}\right)$. At a hydrodynamic level, 
the relevant quantities are the number densities $n_i$, the flow velocity 
${\bf u}$, and the ``granular'' temperature $T$. They are defined in terms 
of moments of the distribution $f_i$ as
\begin{equation}
\label{2.4}
n_i=\int d{\bf v} f_i({\bf v}),  \quad 
\rho{\bf u}=\sum_i\rho_i{\bf u}_i=\sum_i\int d{\bf v}m_i{\bf v}f_i({\bf 
v}),
\end{equation} 
\begin{equation}
\label{2.5}
nT=\sum_in_iT_i=\sum_i\int d{\bf v}\frac{m_i}{d}V^2f_i({\bf v}),
\end{equation} 
where $n=n_1+n_2$ is the total number density, $\rho=\rho_1+\rho_2$ is the 
total mass density, and ${\bf V}={\bf v}-{\bf u}$ is the peculiar 
velocity. Equations (\ref{2.4}) and (\ref{2.5}) also define the flow 
velocity ${\bf u}_i$ and the partial temperature $T_i$ of species $i$, which 
measures the mean kinetic energy of species $i$. 

The collision operators conserve the particle number of each species and the 
total momentum, but the total energy is not conserved. This implies that 
\begin{equation}
\label{2.6}
\sum_{i,j}\int d{\bf v}\case{1}{2}m_{i}V^{2}J_{ij}
[{\bf v}|f_{i},f_{j}]=-\case{d}{2}nT\zeta \;,  
\end{equation}
where $\zeta$ is identified as the ``cooling rate'' due to inelastic
collisions among all species. At a kinetic level, it is also convenient to 
discuss energy transfer in terms of the ``cooling rates'' $\zeta_i$ for the 
partial temperatures $T_i$. They are defined as 
\begin{equation}
\label{2.7}
\zeta_i=-\frac{1}{dn_iT_i}\sum_j\int 
d{\bf v}m_iV^{2}J_{ij}[{\bf v}|f_{i},f_{j}]\;.
\end{equation}
The total cooling rate $\zeta$ can be expressed in terms of $\zeta_i$ as
\begin{equation}
\label{2.9}
\zeta=T^{-1}\sum_ix_iT_i\zeta_i,
\end{equation}
where $x_i=n_i/n$ is the mole fraction of species $i$.

Let us now describe the problem we are interested in. We consider a
granular binary mixture in which the masses
and sizes of both species are arbitrary. Our aim
is to analyze a diffusion problem in a mixture of grains driven into 
flow by the action of shear forces. Here, we consider the
simple case of the tracer limit, namely, a binary mixture in which one of the 
components (say, for instance, 1) is present in tracer concentration 
($x_1\ll 1$). In the tracer limit, one expects that the state of the 
granular gas is not affected by the presence of the 
tracer particles so that its velocity distribution 
function $f_2$ obeys a (closed) nonlinear Boltzmann equation. Furthermore, 
the mole fraction of tracer particles is so small that their mutual 
interactions can be neglected in the kinetic equation of $f_1$. As a 
consequence, the velocity distribution function of tracer particles $f_1$ 
satisfies a (linear) Boltzmann-Lorentz equation.
Let us start describing the state of the mixture in the absence of diffusion.

\subsection{Granular gas (excess component)}

We assume that the granular gas is subjected to the
simple shear flow. From a macroscopic point of view, this state is
characterized by a constant linear velocity profile ${\bf u}={\bf u}_2={\sf 
a}\cdot {\bf r}$, where the elements of the tensor ${\sf a}$ are 
$a_{k\ell}=a\delta_{kx}\delta_{\ell y}$, $a$ being the constant shear rate. 
In addition, the partial density $n\simeq n_2$ and the granular temperature 
$T\simeq T_2$ are uniform. The temporal variation of the granular 
temperature $T_2$ can be obtained from the Boltzmann equation (\ref{2.1}) as 
\begin{equation}
\label{2.10}
\frac{\partial p_2}{\partial t}=-aP_{2,xy}-\frac{d}{2}\zeta_2 p_2,
\end{equation}
where $p_2=n_2T_2$, 
\begin{equation}
\label{2.11}
{\sf P}_2=m_2\int d{\bf v} {\bf V}{\bf V}f_2
\end{equation}
is the pressure tensor of the gas, and in the tracer limit  
\begin{equation}
\label{2.12}
\zeta_2=-\frac{1}{dn_2T_2}\int 
d{\bf v}m_2V^{2}J_{22}[{\bf v}|f_{2},f_{2}]\;.
\end{equation}
The balance of energy (\ref{2.10}) shows the different nature of this state 
for molecular and granular systems. While for elastic fluids ($\zeta_2=0$) 
the temperature increases monotonically in time due to the viscous heating 
term $aP_{2,xy}$, a steady state is possible for granular systems when the 
viscous heating is exactly compensated by the collisional cooling term 
$(d/2)p_2\zeta_2$. In that case, the shear stress $P_{2,xy}$ and the cooling 
rate $\zeta_2$ are related by
\begin{equation}
\label{2.12bis}
aP_{2,xy}=-\frac{d}{2}\zeta_2 p_2.
\end{equation}
As a consequence, for a given shear rate $a$, the (steady) temperature $T_2$ 
is a function of the restitution coefficient $\alpha_{22}$. This steady 
state is what we want to analyze.

The simple shear flow becomes spatially uniform when one refers the 
velocities of the particles to a frame moving with the flow velocity ${\bf 
u}$: $f_2\left({\bf r},{\bf v}\right)\rightarrow f_2({\bf V})$. 
Consequently, the stationary Boltzmann equation for the excess component 
becomes 
\begin{equation}
\label{2.13}
-a V_{y}\frac{\partial}{\partial V_{x}}f_2({\bf V})=J_{22}[{\bf 
V}|f_2,f_2]\;.
\end{equation}
We are mainly interested in computing the nonzero elements of the pressure 
tensor ${\sf P}_2$. These elements can be obtained by 
multiplying the Boltzmann equation (\ref{2.13}) by $m_2 V_{k}V_{\ell}$ and 
integrating over ${\bf V}$. The result is
\begin{equation}
\label{2.14}
a_{k m}P_{2,\ell m}+a_{\ell m}P_{2,km}=A_{k\ell},
\end{equation}
where 
\begin{equation}
\label{2.15}
A_{k\ell}=m_2\int d{\bf V} V_{k}V_{\ell} J_{22}[{\bf V}|f_2,f_2].
\end{equation}
To get an explicit expression for ${\sf P}_2$ one needs to compute the right 
hand side of the set of equations (\ref{2.14}). This requires the explicit 
knowledge of $f_2$, which is not known even in the elastic limit. However, 
one expects to get a good estimate of the low moments of the Boltzmann 
collisional operator by expanding $f_2$ in Sonine polynomials and then
truncate the series after the first few terms. This approach is similar to the 
usual moment method for solving the Boltzmann equation in the elastic case. 
Therefore, we take the leading Sonine approximation:  
\begin{equation}
\label{2.16}
f_2({\bf V})\to f_{2,M}({\bf 
V})\left[1+\frac{m_2}{2T_2}\left(\frac{P_{2,k\ell}}{p_2}-\delta_{k\ell}\right)
\left(V_{k}V_{\ell}-\frac{1}{d}V^2\delta_{k\ell}\right)\right],
\end{equation}
where $f_{2,M}$ is a Maxwellian distribution at the temperature of the 
gas, i.e.,
\begin{equation}
\label{2.17} 
f_{2,M}({\bf V})=n_2 \left(\frac{m_2}{2\pi 
T_2}\right)^{d/2}\exp\left(-\frac{m_2V^2}{2T_2}\right). 
\end{equation} 
With the approximation (\ref{2.16}), the integrals appearing in the 
expressions of the cooling rate $\zeta_2$ and the collisional moment 
$A_{k\ell}$ can be explicitly evaluated. The details of the calculation 
are given in the Appendix. This allows us to get the 
explicit expressions of the non-zero elements of ${\sf P}_2$. They are 
given by  
\begin{equation}
\label{2.18}
P_{2,yy}^*=P_{2,zz}^*=\cdots=P_{2,dd}^*=\frac{d+1+(d-1)\alpha_{22}}
{2d+3-3\alpha_{22}}\;,
\end{equation}
\begin{equation}
\label{2.19}
P_{2,xy}^*=-4d\frac{d+1+(d-1)\alpha_{22}}
{(1+\alpha_{22})(2d+3-3\alpha_{22})^2}a^*\;,
\end{equation}
\begin{equation}
\label{2.20}
P_{2,xx}^*=d-(d-1)P_{2,yy}^*,
\end{equation}
\begin{equation}
\label{2.21}
a^{*2}=\frac{d+2}{32d}
\frac{(1+\alpha_{22})(2d+3-3\alpha_{22})^2(1-\alpha_{22}^2)}{
d+1+(d-1)\alpha_{22}}\;.
\end{equation}
Moreover, the (reduced) cooling rate $\zeta_2^*$ is 
\begin{equation}
\label{2.22}
\zeta_2^*=\frac{d+2}{4d}(1-\alpha_{22}^2).
\end{equation}
Here, ${\sf P}_2^*={\sf P}_2/p_2$, $a^*=a/\nu$, $\zeta_2^*=\zeta_2/\nu$, and 
$\nu$ is a characteristic collision frequency given by $\nu=p_2/\eta$, with 
$\eta$ is the shear viscosity coefficient of the gas in the elastic limit, 
i.e., \begin{equation}
\label{2.23}
\eta=\frac{d+2}{8}\pi^{-(d-1)/2}\Gamma(d/2)
\sigma_{2}^{-(d-1)}(m_2T_2)^{1/2}.
\end{equation}

Equations (\ref{2.18})--(\ref{2.22}) generalize previous results derived in 
the three dimensional case\cite{MG02}. The expression (\ref{2.21}) clearly
indicates the intrinsic connection between the velocity gradient and 
dissipation in the system. As a matter of fact, given that $\alpha_{22}\leq
1$, the range of (reduced) shear rates are defined in the interval $0\leq
a^{*2}\leq (d+2)(3+2d)^2/32d(d+1)$. The parameter $a^*$ can
be considered as the relevant nonequilibrium parameter of the system.
In the elastic limit ($\alpha_{22}=1$ which implies $a^*=0$),
the equilibrium results of the molecular gas are recovered, i.e., 
$P_{2,k\ell}^*=\delta_{k\ell}$. As said in the Introduction, the steady 
simple shear flow for a monocomponent granular fluid has been the subject of 
many previous works. Two interesting studies have been carried out by 
Jenkins and Richman for smooth inelastic disks \cite{JR88} and by Brey {\em 
et al.}\cite{BRM97} for a $d$ dimensional system.
The latter description has been subsequently extended to dense
gases \cite{MGSB99}. The approximated theory of
Jenkins and Richman \cite{JR88} is based in a generalized Maxwellian 
distribution to model the steady shear flow. On the other hand, Brey {\em et 
al.}\cite{BRM97} solved a kinetic model equation of the Boltzmann equation 
and compared their predictions for the reduced temperature, shear stress, 
and normal stress differences with Monte Carlo simulations. Comparison 
between kinetic model results and simulation shows in general a good 
agreement. Our results differ from those reported in Refs.\ \cite{JR88} and 
\cite{BRM97}, although for practical purposes, the discrepancies between the 
different approaches are quite small, even for moderate values of 
$\alpha_{22}$. As an illustration, in Fig.\ \ref{fig1} we compare the 
different theories for the (reduced) elements of the pressure tensor ${\sf 
P}_2^*$ in a two-dimensional system ($d=2$). It is seen that the agreement 
is remarkable, although the discrepancies slightly increase as the 
restitution coefficient decreases.

\subsection{Tracer component}

In the absence of diffusion, the velocity distribution function $f_1$
of tracer particles satisfies the steady kinetic equation
\begin{equation}
\label{3.5}
-a V_{y}\frac{\partial}{\partial V_{x}}f_1=J_{12}[{\bf V}|f_1,f_2]\;.
\end{equation}
The most relevant moment at this level of approximation is the pressure 
tensor ${\sf P}_1$ defined as
\begin{equation}
\label{3.6}
{\sf P}_1=m_1\int d{\bf V} {\bf V}{\bf V}f_1({\bf V}),
\end{equation}
The nonzero elements of ${\sf P}_1$ obey the set of equations:
\begin{equation}
\label{3.7}
a_{k m}P_{1,\ell m}+a_{\ell m}P_{1,km}=B_{k\ell}
\end{equation}
where 
\begin{equation}
\label{3.8}
B_{k\ell}=m_1\int d{\bf V} V_{k}V_{\ell} J_{12}[{\bf V}|f_1,f_2].
\end{equation}
As done before in the case of ${\sf P}_2$, we estimate the 
collisional moment $B_{k\ell}$ by taking the leading Sonine 
approximation of $f_1$ 
\begin{equation}
\label{3.9}
f_1({\bf V})\to f_{1,M}({\bf V})
\left[1+\frac{m_1}{2T_1}\left(\frac{P_{1,k\ell}}{p_1}
-\delta_{k\ell}\right)\left(V_{k}V_{\ell}
-\frac{1}{d}V^2\delta_{k\ell}\right)\right],
\end{equation}
where $p_1=n_1T_1$ and now $f_{1,M}$ is a Maxwellian distribution at the 
temperature of the tracer particles $T_1$, i.e.,
\begin{equation}
\label{3.10} 
f_{1,M}({\bf V})=n_1 \left(\frac{m_1}{2\pi 
T_1}\right)^{d/2}\exp\left(-\frac{m_1V^2}{2T_1}\right). 
\end{equation} 
As will be shown later, the partial temperatures $T_1$ and $T_2$ are in general 
different so that the granular energy per particle is not equally distributed 
between both species.

Once the collisional moment $B_{k\ell}$ is determined (see the 
Appendix), the (reduced) non-zero components of ${\sf P}_1^*=
{\sf P}_1/x_1p_2$ can be easily obtained from Eq.\ (\ref{3.7}). These
components can be written in terms of the temperature ratio $\gamma=T_1/T_2$, 
the restitution coefficients $\alpha_{22}$ and $\alpha_{12}$, and the 
parameters of the mixture. After some algebra, one gets 
\begin{equation}
\label{3.11}
P_{1,yy}^*=P_{1,zz}^*=\cdots=P_{1,dd}^*=-\frac{F+HP_{2,yy}^*}{G},
\end{equation} 
\begin{equation}
\label{3.12}
P_{1,xy}^*=\frac{a^*P_{1,yy}^*-HP_{2,xy}^*}{G},
\end{equation} 
\begin{equation}
\label{3.13}
P_{1,xx}^*=d\gamma-(d-1)P_{1,yy}^*,
\end{equation} 
where 
\begin{equation}
\label{3.14}
F=\frac{\sqrt{2}}{2d}\left(\frac{\sigma_{12}}{\sigma_{22}}
\right)^{d-1}\mu_{12}\left(\frac{1+\theta}{\theta^3}\right)^{1/2}(1+\alpha_{12})
\left[1+\frac{\mu_{21}}{2}(d-1)(1+\theta)(1+\alpha_{12})\right],
\end{equation}
\begin{eqnarray}
\label{3.15}
G&=&-\frac{\sqrt{2}}{4d}\left(\frac{\sigma_{12}}{\sigma_{22}}
\right)^{d-1}\mu_{21}\left(\frac{1}{\theta(1+\theta)}\right)^{1/2}
(1+\alpha_{12})\nonumber\\
& & \times
\left\{2[(d+2)\theta+d+3]-3\mu_{21}
(1+\theta)(1+\alpha_{12})\right\},
\end{eqnarray}
\begin{equation}
\label{3.16}
H=\frac{\sqrt{2}}{4d}\left(\frac{\sigma_{12}}{\sigma_{22}}
\right)^{d-1}\mu_{12}\left(\frac{1}{\theta(1+\theta)}\right)^{1/2}
(1+\alpha_{12})\left[3\mu_{21}(1+\theta)(1+\alpha_{12})-2\right].
\end{equation}
Here, $\theta=m_1T_2/m_2T_1$ is the mean
square velocity of the gas particles relative
to that of the tracer particles. To close the problem at this stage of 
approximation, it still remains to get the temperature 
ratio $\gamma$. It can be obtained for instance, from the requirements 
(\ref{2.12bis}) and its corresponding counterpart for species 1. This yields
\begin{equation}
\label{3.17}
\gamma=\frac{\zeta_2^*P_{1,xy}^*}{\zeta_1^*P_{2,xy}^*},
\end{equation}
where the cooling rate $\zeta_1^*=\zeta_1/\nu$ for the tracer 
particles is (see the Appendix) 
\begin{equation}
\label{3.18}
\zeta_1^*=\frac{(d+2)\sqrt{2}}{4d}\left(\frac{\sigma_{12}}{\sigma_{2}}
\right)^{d-1}\mu_{21}\left(\frac{1+\theta}{\theta}\right)^{1/2}(1+\alpha_{12})
\left[2-\mu_{21}(1+\theta)(1+\alpha_{12})\right].
\end{equation}

The solution to Eq.\ (\ref{3.17}) gives $\gamma$ as a function of the
restitution coefficients $\alpha_{22}$ and $\alpha_{12}$ and
the mechanical parameters of the mixture, i.e., the mass ratio
$\mu=m_1/m_2$ and the size ratio $w=\sigma_1/\sigma_2$.
Except for some limiting cases, Eq.\ (\ref{3.17}) must be solved
numerically. Thus, in the elastic
case ($\alpha_{22}=\alpha_{12}=1$), we recover the well-known equilibrium 
results with $\gamma=1$ and $\theta\to m_1/m_2$ as required by the 
equipartition theorem. In the case of mechanically equivalent particles 
($m_1=m_2$, $\alpha_{22}=\alpha_{12}$, $\sigma_1=\sigma_2$), Eqs.\ 
(\ref{2.18})--(\ref{2.22}) and (\ref{3.11})--(\ref{3.16}) lead to ${\sf 
P}_2^*={\sf P}_1^*$ and $\zeta_2^*=\zeta_1^*$, so that $\gamma=1$. Beyond 
the above limit cases, as expected, our results yield $\gamma \neq 1$. The 
violation of energy equipartition in driven granular mixtures has been 
recently observed in molecular dynamics simulations
of sheared mixtures \cite{CH02} and in real experiments\cite{WP02,FM02}.
This effect is generic of multicomponent
granular systems and is consistent with previous results derived in the 
unforced case\cite{GD99,DG01,SD01,DBL02}. To the best of my knowledge, the only 
previous theories including temperature differences have been proposed by 
Jenkins and Mancini \cite{JM87} and by Huilin {\em et al.} \cite{HGM01}. 
However, both works are phenomenological with no attempt to solve the 
kinetic equation. Instead, they assume that the velocity distribution 
function is a local Maxwellian. This is reasonable for estimating the dense 
gas collisional transfer contributions to the fluxes, but not for evaluating 
their kinetic contributions. Both theories are applicable to a general flow 
field. In particular, the results obtained by Jenkins and Mancini
\cite{JM87} for the temperature ratio in the low density limit for inelastic
disks can be written as 
\begin{equation}
\label{3.18bis}
\gamma=1-\frac{1+\mu}{1+w}\left[\frac{1+w}{2\mu}(1-\alpha_{12})-
\sqrt{\frac{\mu(1+\mu)}{2}}(1-\alpha_{22})\right].
\end{equation} 
In Fig.\ \ref{fig2} we plot the temperature ratio $\gamma$ versus the 
restitution coefficient $\alpha$ for a size ratio $w=2$ and three values of the mass 
ratio $\mu$ in the two-dimensional case ($d=2$). For the sake of 
simplicity, henceforth we will assume that the spheres or disks are made of 
the same material, i.e., $\alpha\equiv \alpha_{22}=\alpha_{12}$. Also for 
comparison, we show the prediction given by Eq.\ (\ref{3.18bis}) in the case 
$\mu=5$. Important discrepancies between both theories appear even for 
values of $\alpha$ close to $1$. As a matter of fact, the theory of Jenkins 
and Mancini predicts a violation of energy equipartition much more 
significant than our theory. It must be remarked that
the quantitative predictions of our theory
at the level of the temperature ratio have been recently confirmed by Monte 
Carlo simulations\cite{MG02,GM02}. Regarding the influence 
of the parameters of the mixture, we observe that for large mass ratios
the temperature differences are quite significant, even for
moderate dissipation (say $\alpha \simeq 0.9$). The temperature of the 
tracer particles is larger than that of the excess species when the tracer 
grains are heavier than the grains of the gas.
This behavior has been also found in the recent computer simulations
carried out in rapid shear flow \cite{CH02}.

\section{Tracer diffusion under simple shear flow} 
\label{sec3}

We want to study the diffusion of tracer particles immersed in a bath 
(granular gas)  subjected to the simple shear flow. The diffusion process is 
induced in the system by a {\em weak} concentration gradient $\nabla x_1$. 
However, given that the strength of the shear rate $a$ is arbitrary,
the mass flux (which is generated by the gradient $\nabla x_1$) can be modified by the presence 
of the shear flow. As stated above, in the tracer limit the state of 
particles of species 1 is mainly governed by the collisions with particles 
of species 2, so that the self-collisions among particles 1 can be 
neglected. Thus, the kinetic equation governing the evolution 
of the velocity distribution function $f_1$ reads
\begin{equation}
\label{3.1}
\frac{\partial}{\partial t}f_1-a V_{y}\frac{\partial}{\partial 
V_{x}}f_1+(V_k+a_{k\ell}r_{\ell})\frac{\partial}{\partial 
r_k}f_1=J_{12}[{\bf V}|f_1,f_2]\;,
\end{equation}
where here the derivative $\partial/\partial r_k$ is
taken at constant ${\bf V}$.
Tracer particles may freely exchange momentum and energy with the particles
of the granular gas and, therefore, these are not invariants of the 
collision operator $J_{12}[f_1,f_2]$. Only the number density of tracer 
particles is conserved. More specifically, the mole fraction $x_1$ obeys the 
conservation law
\begin{equation}
\label{3.2}
\left(\frac{\partial}{\partial t}+a_{k\ell}r_\ell\frac{\partial}{\partial 
r_k}\right)x_1+\frac{\nabla\cdot {\bf j}_1}{m_1}=0,
\end{equation}
where the mass flux ${\bf j}_1$ is defined as 
\begin{equation}
\label{3.3}
{\bf j}_1=m_1\int\, d{\bf V} {\bf V} f_1({\bf V}).
\end{equation}

When $\alpha_{22}=\alpha_{12}=1$ (which gives $a^*=0$),
the well-known Fick law establishes a linear relationship between the mass flux
${\bf j}_1$ and the concentration
gradient $\nabla x_1$. This law defines the diffusion coefficient. 
For finite values of the (reduced) shear rate $a/\nu$ (which means 
$\alpha_{22}\neq 1$), one expects that a generalized
Ficks's law holds but now a diffusion tensor rather than a scalar should 
appear. Our aim is to get this tensor in terms of $\alpha_{22}$,
$\alpha_{12}$, $\mu$, and $w$. To this end, and assuming that the mole fraction
$x_1$ is slightly {\em nonuniform}, we solve Eq.\ (\ref{3.1}) by means of a 
perturbation expansion around a nonequilibrium state with arbitrary shear 
rate, which is equivalent to strong dissipation in the simple shear flow 
[see Eq.\ (\ref{2.21})]. Thus, we write
\begin{equation}
\label{3.4}
f_1=f_1^{(0)}+f_1^{(1)}+\ldots, 
\end{equation}
where $f_1^{(k)}$ is of order $k$ in $\nabla x_1$ but applies for {\em 
arbitrary} degree of dissipation since this distribution retains all the 
orders in $a$. The solution (\ref{3.4}) qualifies as a normal solution since
all the space and time dependence of $f_1$ occurs entirely
through $x_1({\bf r};t)$ and their gradients. The zeroth-order approximation $f_1^{(0)}$ corresponds to the
simple shear flow distribution but taking into account now the local 
dependence on the mole fraction $x_1$. Although the explicit form of 
$f_1^{(0)}$ is not exactly known, only the knowledge of its second-degree 
moments (related to the pressure tensor ${\sf P}_1$) is necessary to get the
diffusion tensor in the Sonine approximation.

The kinetic equation for $f_1^{(1)}$ can be obtained from the Boltzmann-Lorentz
equation (\ref{3.1}) by collecting all the terms of
first order in $\nabla x_1$: 
\begin{equation}
\label{3.19}
\frac{\partial}{\partial t}f_1^{(0)}-a V_{y}\frac{\partial}{\partial 
V_{x}}f_1^{(1)}+(V_i+a_{k\ell}r_\ell)\frac{\partial}{\partial 
r_k}f_1^{(0)}=J_{12}[{\bf V}|f_1^{(1)},f_2]\;.
\end{equation} 
According to the balance equation (\ref{3.2}), one has 
\begin{equation}
\label{3.20}
\frac{\partial f_1^{(0)}}{\partial t}=\frac{\partial f_1^{(0)}}{\partial x_1} 
\frac{\partial x_1}{\partial t}=-a_{k\ell}r_\ell\frac{\partial 
x_1}{\partial r_k}\frac{\partial f_1^{(0)}}{\partial x_1}
\end{equation}
where use has been made of the fact that the zeroth-order approximation 
to the mass flux vanishes, i.e., ${\bf j}_1^{(0)}={\bf 0}$. Moreover, 
\begin{equation}
\label{3.20bis}
\frac{\partial f_1^{(0)}}{\partial r_k}=
\frac{\partial f_1^{(0)}}{\partial x_1}\frac{\partial x_1}{\partial 
r_k}
\end{equation}
Using (\ref{3.20}) and (\ref{3.20bis}), Eq.\ (\ref{3.19}) can be written as 
\begin{equation}
\label{3.21}
\left(a V_{y}\frac{\partial}{\partial V_{x}}+\Lambda\right)f_1^{(1)}=
\frac{\partial f_1^{(0)}}{\partial x_1}\left({\bf V}\cdot \nabla 
x_1\right)\;,
\end{equation}
where $\Lambda$ is the Boltzmann-Lorentz collision operator 
\begin{equation}
\label{3.22}
\Lambda f_1^{(1)}=J_{12}[{\bf V}|f_1^{(1)},f_2].
\end{equation}
The solution to Eq.\ (\ref{3.21}) is proportional to $\nabla x_1$, i.e., it 
has the form
\begin{equation}
\label{3.23}
f_1^{(1)}({\bf V})={\sf {\cal A}}({\bf V})\cdot \nabla x_1
\end{equation}
Substitution of this into Eq.\ (\ref{3.21}) yields 
\begin{equation}
\label{3.24}
\left(a V_{y}\frac{\partial}{\partial V_{x}}+\Lambda\right){\sf {\cal A}}=
\frac{\partial f_1^{(0)}}{\partial x_1}{\bf V}\;.
\end{equation} 
The first order approximation to the mass flux is given by  
\begin{eqnarray}
\label{3.25}
{\bf j}_1^{(1)}&=&m_1\int\, d{\bf V} {\bf V} f_1^{(1)}({\bf V}) \nonumber\\
&=& -{\sf D}\cdot \nabla x_1,
\end{eqnarray}
where the tracer diffusion tensor is
\begin{equation}
\label{3.26}
D_{k\ell}=-m_1\int\, d{\bf V} V_k\, {\cal A}_\ell({\bf V}).
\end{equation}
The solution to the integral equation (\ref{3.24}) allows one to determine 
the quantity ${\sf {\cal A}}$. From this solution one can determine
the tracer diffusion tensor by means of Eq.\ (\ref{3.26}).

In order to get an explicit expression for the tensor ${\sf D}$ we need to 
know the quantity ${\sf {\cal A}}$. A good estimate of ${\sf {\cal A}}$ to 
evaluate the mass flux ${\bf j}_1^{(1)}$ is given by the first Sonine 
approximation, in which only the leading term in the expansion of ${\sf 
{\cal A}}({\bf V})$ in Sonine polynomials is kept. Thus, we take the 
following approximation to ${\sf {\cal A}}$: 
\begin{equation}
\label{3.27}
{\sf {\cal A}}({\bf V})\to-\frac{1}{n_1T_1}{\bf V}\cdot {\sf D}f_{1,M}({\bf 
V}).
\end{equation}
Using Eq.\ (\ref{3.27}), an equation for the diffusion tensor is easily 
derived from Eq.\ (\ref{3.24}). The result can be written as 
\begin{equation}
\label{3.28}
\left({\sf a}+{\bbox {\Omega}}\right)\cdot {\sf 
D}=p_2{\sf P}_{1}^{*},
\end{equation}
where the nonzero components of ${\sf P}_1^{*}$ are given by Eqs.\ 
(\ref{3.11})--(\ref{3.13}) and we have introduced the tensorial quantity
\begin{equation}
\label{3.29}
{\bbox {\Omega}}=-\frac{m_1}{n_1T_1}\int\,d{\bf V}\,{\bf V}\,\Lambda 
{\bf V} f_{1,M}.
\end{equation}  
The expresion of the tensor ${\bbox {\Omega}}$ has been obtained in the 
Appendix with the result
\begin{equation}
\label{3.30}
{\bbox {\Omega}}=\frac{2}{d}\frac{\pi^{(d-1)/2}}{\Gamma(d/2)}
n_2\mu_{21}\sigma_{12}^{d-1}(2T_2/m_2)^{1/2}
(1+\alpha_{12})\left[(1+\theta)\theta\right]^{-1/2}
\left[\left(1+\frac{d+1}{d+2}\theta\right)\openone
+\frac{\theta}{d+2}{\sf P}_{2}^*\right].
\end{equation}

The solution of Eq.\ (\ref{3.28}) is 
\begin{equation}
\label{3.31}
{\sf D}=p_2\left({\sf a}+{\bbox {\Omega}}\right)^{-1}\cdot {\sf P}_1^{*}
\end{equation}
Equation (\ref{3.31}) is the primary result of this paper. It 
provides an explicit expression of the tracer diffusion tensor of a {\em 
granular} binary mixture in simple shear flow. The elements of this tensor 
give all the information on the physical mechanisms involved in the 
diffusion of tracer particles in a strongly sheared granular gas. 
In the absence of shear rate (what is equivalent to 
$\alpha_{22}=\alpha_{12}=1$), $D_{k\ell}=D_0\delta_{k\ell}$, where 
\begin{equation}
\label{3.32}
D_0=\frac{d}{4\sqrt{2}}\frac{\Gamma(d/2)}{\pi^{(d-1)/2}
\sigma_{12}^{d-1}}\sqrt{\mu(1+\mu)}\left(m_2T_2\right)^{1/2}
\end{equation}
is the tracer diffusion coefficient of a molecular gas\cite{CC70}. As the 
restitution coefficients decrease, rheological effects become important and 
the elements of the diffusion tensor are different from the one obtained in 
the equilibrium case. The dependence of the diffusion coefficients 
on the restitution coefficients $\alpha_{22}$ and $\alpha_{12}$ as well as 
on the mass ratio $\mu$ and the size ratio $w$ is highly nonlinear. 
As happens for elastic fluids\cite{G}, Eq.\ (\ref{3.31}) shows that diffusion 
under simple shear flow is a very complex problem due basically to the 
anisotropy induced in the system by the shear flow.

To illustrate the dependence of the elements $D_{k\ell}$ on the parameters 
of the problem, let us consider a mixture of inelastic hard-spheres ($d=3$). 
According to Eq.\ (\ref{3.31}), 
$D_{xz}=D_{zx}=D_{yz}=D_{zy}=0$, in agreement with the symmetry of the 
problem. Consequently, there are five relevant elements: the three diagonal 
and two ($D_{xy}$, $D_{yx}$) off-diagonal elements. In general,
$D_{xx}\neq D_{yy}\neq D_{zz}$ and $D_{xy}\neq D_{yx}$.
The off-diagonal elements measure cross effects
in the diffusion of particles induced by the shear flow. Thus, for instance 
$D_{xy}$ gives the transport of mass along the direction of the flow of 
the system ($x$ axis) due to a concentration gradient parallel to the 
gradient of the flow velocity ($y$ axis). Both off-diagonal elements are negative.

Before analyzing the influence of the mechanical parameters of the mixture 
on diffusion, it is instructive to explore the particular case of 
self-diffusion, i.e., when the tracer particles are mechanically equivalent 
to the gas particles. This situation involves only 
single-particle motion and it is therefore somewhat simpler to compute the 
diffusion coefficients. In particular, the temperature of the tracer 
particles is the same as that of the gas particles and so $\theta=1$ in 
Eqs.\ (\ref{3.11})--(\ref{3.16}) and (\ref{3.30}). In Fig.\ \ref{fig3}, we 
plot $D_{xx}^*-D_{yy}^*$, $D_{zz}^*-D_{yy}^*$, 
$(D_{xx}^*+D_{yy}^*+D_{zz}^*)/3\equiv (1/3)D_{kk}^*$, $-D_{xy}^*$, and 
$-D_{yx}^*$ as functions of the restitution coefficient $\alpha\equiv 
\alpha_{12}=\alpha_{22}$. Here, $D_{ij}^*\equiv D_{ij}/D_0$, with $D_0$ 
given by Eq.\ (\ref{3.32}). We see that the deviation from the functional 
form for elastic collisions is quite important even for moderate 
dissipation. Thus, for instance at $\alpha=0.8$, $D_{xx}^*-D_{yy}^*\simeq 
0.76$, $D_{zz}^*-D_{yy}^*\simeq 0.046$, $(1/3)D_{kk}^*\simeq 1.18$, 
$-D_{xy}^*\simeq 1.039$, and $-D_{yx}^*\simeq 0.42$. The figure also shows 
that the anisotropy of the system, as measured by the differences 
$D_{xx}^*-D_{yy}^*$ and  $D_{zz}^*-D_{yy}^*$, grows with the inelasticity. 
This anisotropy is much more important in the plane of shear flow 
($D_{xx}^*-D_{yy}^*$) that in the plane perpendicular to the flow velocity 
($D_{zz}^*-D_{yy}^*$). This is basically due to the fact that 
$P_{s,xx}^*\neq P_{s,yy}^*=P_{s,zz}^*$ with $s=1,2$.

As said before, Campbell \cite{C97} has carried out molecular dynamics 
simulations to measure the nonzero elements of the self-diffusion tensor. 
In his work, the self-diffusion coefficients were nondimensionalized by the 
product of the shear rate and the particle diameter. In our units, this 
corresponds to the reduced tensor $\widetilde{D}_{ij}=D_{ij}^*/a^*$. 
Although the solid fractions analyzed in his simulations prevent us in 
general from making a quantitative comparison between our theory (restricted to 
dilute gases) and his computer simulations, we observe that the 
general qualitative dependence of the self-diffusion tensor on dissipation 
agrees well with our results, at least for the lowest solid fraction 
considered. Thus, theory and simulation predict 
that the magnitude of the normal diffusion coefficients follow the pattern 
$\widetilde{D}_{xx}>\widetilde{D}_{zz}>\widetilde{D}_{yy}$ while, in general, 
the elements $\widetilde{D}_{ij}$ decrease as the restitution coefficient 
decrease. An exception to the latter rule is the element 
$\widetilde{D}_{xx}$, which does not depend sensitively on $\alpha$.
On the other hand, in Campbell's simulation work \cite{C97}, he found that 
the values of $-\widetilde{D}_{xy}$ were roughly of the same magnitude as 
$\widetilde{D}_{yy}$ provided that the solid fraction is smaller than 0.4. 
This trend is not completely followed by our theory since the 
values of $-\widetilde{D}_{xy}$ and $\widetilde{D}_{yy}$
are significantly different for highly inelastic spheres. Thus, for
instance, at $\alpha=0.8$, $-\widetilde{D}_{xy}\simeq 2.43$ and 
$\widetilde{D}_{yy}\simeq 2.15$ but  $-\widetilde{D}_{xy}\simeq 2.54$ and 
$\widetilde{D}_{yy}\simeq 1.10$ at $\alpha=0.4$.

The dependence of the diffusion coefficients $D_{ij}^*$ on the 
restitution coefficient for different values of the mass ratio $\mu$
is illustrated in Fig.\ \ref{fig4}. In this case,we take a size ratio $w=2$
and two values of the mass ratio: $\mu=2$ and
$\mu=4$. For a given value of the inelasticity, we observe that the deviations 
from the elastic results are more important as the tracer particles are heavier 
than the gas particles.

\section{Concluding remarks}
\label{sec4}

In this paper, we have described diffusion of tracer particles in a granular 
gas subjected to the simple shear flow. We have been interested in the steady 
state where the effect of viscosity is compensated for by the dissipation in 
collisions. Under these conditions, the resulting diffusion is anisotropic 
and, thus, cannot be described by a single diffusion coefficient. Instead, 
it must be described by a diffusion tensor whose explicit determination has 
been the main objective of this work. In order to capture the essential 
aspects of such a nonlinear problem, we have considered here a granular 
mixture in the low density regime as a prototype granular system, which 
lends itself to a detailed description by means of the nonlinear Boltzmann 
kinetic equation.

We have been concerned with the physical situation where a weak 
concentration gradient coexists with a strong shear rate, which for the 
steady simple shear flow problem means strong dissipation. For this reason, 
the Boltzmann-Lorentz equation corresponding to the tracer particles 
has been solved by means of a perturbation expansion around a {\em nonequilibrium} 
sheared state. This implies that the different approximations of this 
expansion are {\em nonlinear} functions of the restitution coefficients as 
well as of the parameters of the mixture (mass and size ratios).  To get 
explicit results, we have used a first Sonine polynomial approximation to 
evaluate the cooling rates and the collisional moments of the Boltzmann 
operators. The reliablity of this approximation has been recently assessed  
in the (pure) shear flow problem where it has been shown to agree very 
well with Monte Carlo simulations in the case of hard spheres \cite{MG02}.

The kinetic theory results show that the elements $D_{ij}$ of the diffusion 
tensor present a complex dependence on the restitution coefficients 
$\alpha_{22}$ and $\alpha_{12}$ and on the mass ratio $\mu=m_1/m_2$ and the 
size ratio $w=\sigma_1/\sigma_2$. In the elastic case, $\alpha_{ij}=1$, 
$D_{ij}=D_0 \delta_{ij}$, where $D_0$ is given by Eq.\ (\ref{3.32}) and one 
recovers the expression of the diffusion coefficient for normal fluids. 
The deviations of the tensor $D_{ij}$ from the scalar $D_0$ have two
distinct origins.  First, the presence of shear flow gives rise to the 
new tensorial term ${\sf a}+{\bbox {\Omega}}$ on the left hand side of Eq.\ 
(\ref{3.28}) instead of the corresponding collision frequency of the elastic 
diffusion problem. Second, given that the tracer and fluid particles are 
mechanically different, the reference state (zeroth-order approximation of 
the expansion) of tracer particles is completely different from that of the 
gas particles. In particular, when $\mu\neq 1$ and/or $w\neq 1$, the 
temperature ratio $\gamma$ is clearly different from $1$
(as can be seen in Fig.\ \ref{fig2}), confirming the breakdown of
the energy equipartition. The
effect of different temperatures for the tracer and gas particles is 
expressed by the appearance of $\theta=\mu/\gamma$ in ${\sf P}_1$
[cf. Eqs.\ (\ref{3.11})--(\ref{3.13})] and in ${\bbox {\Omega}}$.
Each one of the two
afforementioned effects is a different reflection of dissipation present in 
the system.

A simple case is the self-diffusion problem, i.e., when the tracer and 
gas particles are mechanically equivalent. In this case, ${\sf 
P}_2^*={\sf P}_1^*$ and $\gamma=1$. This situation 
has been previously studied by Campbell \cite{C97} by means of molecular 
dynamics simulations. As has been discussed in Sec.\ \ref{sec3}, our 
predictions for the self-diffusion tensor agree qualitatively well with 
these simulations. On the other hand, when the tracer and gas particles are 
mechanically {\em different}, to my knowledge no previous studies on the 
diffusion tensor under shear flow have been made. As pointed out in the 
Introduction, most of the works on granular mixtures \cite{JM89} are based 
on the Chapman-Enskog expansion around a local equilibrium state up to the 
Navier-Stokes order, and therefore, they are restricted to the 
low-dissipation limit in the simple shear flow. In addition, they also 
assume a single temperature to describe the mixture. A more careful 
calculation which takes into account temperature differences has been 
recently made by Garz\'o and Dufty\cite{GD02,DG01}. They have obtained explicit 
expressions for the transport coefficients of a granular binary mixture in 
terms of the restitution coefficients and the parameters of the mixture. 
Since these results have been derived taking the freely cooling state as the 
reference one, the diffusion is characterized by a single scalar
coefficient that cannot be directly compared to the diffusion
tensor obtained here.
However, it would be interesting to compare the diffusion results obtained 
here in the driven sheared case with those found in the unforced 
case\cite{GD02}. In Fig.\ \ref{fig5}, we have compared the behavior of the 
scalar $\case{1}{3}D_{kk}^*$ (which can be understood as a generalized 
mutual diffusion coefficient in a sheared mixture) with the (reduced) 
diffusion coefficient $D^*$ obtained in Ref.\ \onlinecite{GD02} in the
tracer limit ($x_1\to 0$).
We observe that, although the reference states in
both descriptions are very different, the dependence of both diffusion 
coefficients on dissipation is quite similar since they increase as 
$\alpha$ increases. This trend is more significant in the unforced case than 
in the sheared case.   
 
The evaluation of the diffusion tensor for practical purposes 
requires the truncation of a Sonine polynomial expansion. In the case of the 
(pure) simple shear flow problem, recent Monte Carlo 
simulations \cite{MG02,GM02} have shown the accuracy of the leading order 
truncation. We expect that 
this agreement may be extended to the elements of the diffusion tensor for a 
wide range of values of dissipation. 
Exceptions to this agreement
could be disparate mass binary mixtures (e.g., electron-proton systems) for
which the first Sonine solution is not perhaps a good approximation and 
higher-order terms should
be considered. We hope that the results derived in this 
paper will stimulate the performance of
computer simulations to check the quality of the approximations given here 
for the diffusion tensor. Given the difficulties associated with molecular 
dynamics simulations in the low-density regime, one could perhaps use the 
direct simulation Monte Carlo method \cite{B94} which is being shown to be 
fruitful in the context of granular systems.

\acknowledgments

Partial support from the Ministerio de Ciencia y Tecnolog\'{\i}a (Spain)
through Grant No.\ BFM2001-0718 is ackowledged. The author is grateful
to Andr\'es Santos for suggestions for the improvement of the
manuscript.

\appendix

\section{Evaluation of \lowercase{$\zeta_i$}, ${\sf A}$, ${\sf B}$, and 
${\bf \Omega}$}

In this Appendix we evaluate the cooling rates $\zeta_i$, the collisional 
velocity moments ${\sf A}$, and ${\sf B}$, and the tensorial collision 
frequency ${\bbox {\Omega}}$ by using the corresponding leading Sonine 
approximations.  

\subsection{Evaluation of $\zeta_i$} 

The cooling rate $\zeta_1$ of the tracer particles is defined as  
\begin{equation}
\label{a1}
\zeta_1=-\frac{1}{dn_1T_1}\int 
d{\bf V}_{1}m_{1}V_{1}^{2}J_{12}[{\bf V}_{1}|f_{1},f_{2}]\;.
\end{equation}
A useful identity for an arbitrary function $h({\bf V}_{1})$ is given by  
\begin{eqnarray} 
\label{a2}
\int d{\bf V}_{1}h({\bf V}_{1})J_{12}\left[ {\bf 
V}_{1}|f_{1}^{(0)},f_{2}\right] &=&\sigma _{12}^{d-1}\int \,d{\bf 
V}_{1}\,\int \,d{\bf V}_{2}f_{1}^{(0)}({\bf V}_{1})f_{2}({\bf V}_{2})  
\nonumber\\
& & \times \int d\widehat{\bbox {\sigma}}\,\Theta (\widehat{\bbox 
{\sigma}} \cdot {\bf g})(\widehat{\bbox {\sigma}}\cdot {\bf g})
\left[h({\bf V}_1'')-h({\bf V}_1)\right],
\end{eqnarray} 
with  
\begin{equation}
\label{a3} 
{\bf V}_{1}^{^{\prime \prime}}={\bf V}_{1}-\mu _{21}(1+\alpha _{12})(  
\widehat{\bbox {\sigma }}\cdot {\bf g})\widehat{\bbox {\sigma}}\;. 
\end{equation} 
Using (\ref{a2}), Eq.\ (\ref{a1}) can be written as
\begin{eqnarray}
\label{a4} 
\zeta_1 
&=&-\frac{m_{1}}{dn_{1}T_{1}}\sigma_{12}^{d-1}(1+\alpha_{12})\mu_{21}
\int \,d{\bf V}_{1}\,\int \,d{\bf V}_{2}f_{1}({\bf V}_{1})f_{2}({\bf 
V}_{2})\int d\widehat{\bbox {\sigma}}\,\Theta (\widehat{\bbox 
{\sigma}} \cdot {\bf g}_{12})(\widehat{\bbox {\sigma}}\cdot {\bf g}_{12})^2
\nonumber\\
& & \times\left[\mu_{21}(1+\alpha_{12})
(\widehat{\bbox {\sigma}}\cdot {\bf g}_{12})-2({\bf V}_1\cdot 
\widehat{\bbox {\sigma}})\right]
\end{eqnarray} 
To perform the angular integrations, we need the results 
\begin{equation}
\label{a5}
\int d\widehat{\bbox {\sigma}}\,\Theta (\widehat{\bbox 
{\sigma}} \cdot {\bf g}_{12})(\widehat{\bbox {\sigma}}\cdot {\bf 
g}_{12})^n=\beta_n g_{12}^n,
\end{equation}
\begin{equation}
\label{a6}
\int d\widehat{\bbox {\sigma}}\,\Theta (\widehat{\bbox 
{\sigma}} \cdot {\bf g}_{12})(\widehat{\bbox {\sigma}}\cdot {\bf g}_{12})^n
\widehat{\bbox {\sigma}}=\beta_{n+1} g_{12}^{n-1}{\bf g}_{12},
\end{equation}
where 
\begin{equation}
\label{a7}
\beta_n=\pi^{(d-1)/2}\frac{\Gamma\left((n+1)/2\right)}
{\Gamma\left((n+d)/2\right)}.
\end{equation}
Thus, the integration over $\widehat{\bbox {\sigma}}$ in Eq.\ (\ref{a4}) 
leads to 
\begin{eqnarray}
\label{a8}
\zeta_1&=&\frac{m_{1}}{dn_{1}T_{1}}\sigma_{12}^{d-1}\beta_3
(1+\alpha_{12})\mu_{21}
\int \,d{\bf V}_{1}\,\int \,d{\bf V}_{2}f_{1}({\bf V}_{1})f_{2}({\bf 
V}_{2})\nonumber\\
& & \times \left[\mu_{21}g_{12}^3(1-\alpha_{12})+2g_{12}({\bf g}_{12}\cdot 
{\bf G}_{12})\right],
\end{eqnarray}
where ${\bf G}_{12}=\mu_{12}{\bf V}_1+\mu_{21}{\bf V}_2$ is the center of 
mass velocity. Now, we take the Sonine approximations given by (\ref{2.16}) 
for $f_2$ and (\ref{3.9}) for $f_1$. Neglecting nonlinear terms in the 
tensors ${\sf C}_i\equiv {\sf P}_{i}^*-\openone$, one gets
\begin{equation}
\label{a9} 
\zeta_1=\frac{m_1n_{2}}{dT_1}\sigma_{12}^{d-1}v_0^3\beta_3
\mu_{21}(1+\alpha_{12})\theta^{d/2}I_{\zeta}(\theta),
\end{equation}
where 
\begin{equation}
\label{a10}
I_{\zeta}(\theta)=\pi^{-d}\int \,d{\bf V}_{1}^{\ast}\,\int 
\,d{\bf V}_{2}^{\ast}g_{12}^{\ast} 
\left[\mu_{21}g_{12}^{*2}(1-\alpha_{12})+2({\bf g}_{12}^*\cdot {\bf 
G}_{12}^*)\right]
e^{-\theta V_{1}^{\ast 2}-V_{2}^{\ast 2}}.
\end{equation} 
Here, ${\bf V}_{i}^{\ast}={\bf V}_{i}/v_{0}$, ${\bf g}_{12}^{\ast }={\bf 
g}_{12}/v_{0}$, ${\bf G}_{12}^{\ast}={\bf 
G}_{12}/v_{0}$, $\theta=\mu_{12}/(\mu_{21}\gamma)$, $\gamma=T_1/T_2$, and 
$v_{0}=\sqrt{2T_{2}/m_{2}}$. In Eqs.\ (\ref{a9}) and (\ref{a10}), use has 
been made of the fact that the scalar $\zeta_1$ cannot be coupled to the 
traceless tensor ${\sf C}_i$ so that the only contributions to $\zeta_1$ 
come from the (pure) Maxwellian terms in (\ref{2.16}) and (\ref{3.9}). The 
integral $I_{\zeta}$ can be evaluated by the change of variables
\begin{equation}
\label{a11} 
{\bf x}={\bf V}_{1}^{\ast}-{\bf V}_{2}^{\ast},\quad {\bf y}=\theta  
{\bf V}_{1}^{\ast}+{\bf V}_{2}^{\ast},   
\end{equation} 
with the Jacobian $\left(1+\theta\right)^{-d}$. The integral 
$I_{\zeta}(\theta)$ can be now easily computed with the result
\begin{equation}
\label{a12}
I_{\zeta}(\theta)=\frac{\Gamma\left((d+3)/2\right)}
{\Gamma\left(d/2\right)}
\theta^{-(d+3)/2}(1+\theta)^{1/2}\left[2-\mu_{21}(1+\theta)
(1+\alpha_{12})\right].
\end{equation}
Use of the result (\ref{a12}) in Eq.\ (\ref{a9}) yields 
\begin{equation}
\label{a13}
\zeta_1=2\frac{\pi^{(d-1)/2}}{d\Gamma\left(d/2\right)}
n_2\sigma_{12}^{d-1}v_0\mu_{21}(1+\alpha_{12})
\left(\frac{1+\theta}{\theta}\right)^{1/2}\left[2-\mu_{21}(1+\theta)
(1+\alpha_{12})\right].
\end{equation}
The corresponding expression for $\zeta_2$ can be easily obtained from 
(\ref{a13}) and the result is
\begin{equation}
\label{a14}
\zeta_2=\sqrt{2}\frac{\pi^{(d-1)/2}}{d\Gamma(d/2)}
n_2\sigma_{2}^{d-1}v_0(1-\alpha_{22}^2).
\end{equation}

\subsection{Evaluation of ${\sf A}$ and ${\sf B}$} 

Since the tensor ${\sf A}$ can be easily obtained from the expression of 
${\sf B}$, let us explicitly evaluate the latter tensor. It is defined as 
\begin{equation}
\label{b1}
{\sf B}=\int\,d{\bf V}_1 m_1{\bf V}_{1}{\bf V}_{1}J_{12}[{\bf 
V}_1|f_1,f_2]
\end{equation}
Using (\ref{a2}), Eq.\ (\ref{b1}) can be written as
\begin{eqnarray}
\label{b2}
{\sf B}&=&-m_1\sigma_{12}^{d-1}\beta_3\mu_{21}(1+\alpha_{12})
\int \,d{\bf V}_{1}\,\int \,d{\bf V}_{2}f_{1}({\bf V}_{1})f_{2}({\bf 
V}_{2})g_{12}\nonumber\\
& & \times 
\left[{\bf g}_{12}{\bf G}_{12}+{\bf G}_{12}{\bf g}_{12}
+\frac{\mu_{21}}{d+3}(2d+3-3\alpha_{12}){\bf g}_{12}{\bf g}_{12}
-\frac{\mu_{21}}{d+3}
(1+\alpha_{12})g_{12}^2\openone\right],
\nonumber\\
\end{eqnarray}
where use has been made of (\ref{a6}) and 
\begin{equation}
\label{b3}
\int d\widehat{\bbox {\sigma}}\,\Theta (\widehat{\bbox 
{\sigma}} \cdot {\bf g}_{12})(\widehat{\bbox {\sigma}}\cdot {\bf g}_{12})^n
\widehat{\bbox {\sigma}}\widehat{\bbox {\sigma}}
=\frac{\beta_{n}}{n+d} g_{12}^{n-2}\left(n{\bf g}_{12}{\bf g}_{12}+g_{12}^2
\openone\right).
\end{equation}
Substituting the Sonine approximations (\ref{2.16}) and (\ref{3.9}) 
for $f_2$ and $f_1$, respectively, and retaining only linear terms in 
the tensors ${\sf C}_i$, one gets
\begin{eqnarray}
\label{b4}
{\sf B}&=&-m_1\sigma_{12}^{d-1}\beta_3\mu_{21}(1+\alpha_{12})
n_1n_2v_0^3\theta^{d/2}\pi^{-d}\int \,d{\bf V}_{1}^{\ast}\,\int 
\,d{\bf V}_{2}^{\ast}g_{12}^{\ast} 
e^{-\theta V_{1}^{\ast 2}-V_{2}^{\ast 2}}\nonumber\\ 
& &\times  
\left[1+\theta{\sf C}_1: \left({\bf V}_{1}^*{\bf 
V}_{1}^*-\case{1}{d}V_1^{*2}\openone\right)+
{\sf C}_2:\left({\bf V}_{2}^*{\bf 
V}_{2}^*-\case{1}{d}V_2^{*2}\openone\right)\right]\nonumber\\
& & \times  
\left[{\bf g}_{12}^*{\bf G}_{12}^*+{\bf G}_{12}^*{\bf g}_{12}^*
+\frac{\mu_{21}}{d+3}
(2d+3-3\alpha_{12}){\bf g}_{12}^*{\bf g}_{12}^*-\frac{\mu_{21}}{d+3}
(1+\alpha_{12})g_{12}^{*2}\openone\right].
\end{eqnarray}
This integral can be computed by the change of variables
(\ref{a11}). After a lengthy algebra, the result is 
\begin{eqnarray}
\label{b6}
{\sf B}&=&\frac{\pi^{(d-1)/2}}{d\Gamma(d/2)}\sigma_{12}^{d-1}m_1
n_1n_2\mu_{21}(1+\alpha_{12})v_0^3\left(\frac{1+\theta}{\theta}\right)^{3/2}
\nonumber\\
& & \times\left\{\left[\frac{\lambda_{12}}{d+2}+\frac{d}{d+3}\mu_{21}
(1+\alpha_{12})\right]\openone-2\frac{\theta}{(1+\theta)^2}\left[
\left(1+\frac{(d+3)}{2(d+2)}\frac{1+\theta}{\theta}\lambda_{12}\right)
\gamma^{-1}{\sf P}_{1}^*\right.\right. \nonumber\\
& & \left.\left. 
-\left(1-\frac{(d+3)}{2(d+2)}(1+\theta)\lambda_{12}\right){\sf P}_{2}^*
\right]\right\},
\end{eqnarray}
where 
\begin{equation}
\label{b7}
\lambda_{12}=\frac{2}{1+\theta}-\frac{3}{d+3}\mu_{21}(1+\alpha_{12}).
\end{equation}

The corresponding expression for ${\sf A}$ can be easily 
inferred from Eq.\ (\ref{b6}) by just making the change $1\to 2$ and 
$\theta\to 1$:  
\begin{equation}
\label{b8}
{\sf A}=\frac{\sqrt{2}\pi^{(d-1)/2}}{2d(d+2)\Gamma(d/2)}
\sigma_{2}^{d-1}m_2n_2^2v_0^3(1+\alpha_{22})\left\{[d+1+\alpha_{22}(d-1)]
\openone-(2d+3-3\alpha_{22}){\sf P}_{2}^*\right\}.
\nonumber\\
\end{equation}

\subsection{Evaluation of ${\bf \Omega}$} 

The tensor ${\bbox {\Omega}}$ is defined as
\begin{equation}
\label{c1}
{\bbox {\Omega}}=-\frac{m_1}{n_1T_1}\int\,d{\bf V}_1\,
{\bf V}_1\,J_{12}[{\bf V}_1 f_{1,M},f_2].
\end{equation}
The evaluation of ${\bbox {\Omega}}$ can be made following similar 
mathematical steps as above. Thus, using (\ref{a1}) and the Sonine 
approximation to $f_2$ in Eq.\ (\ref{c1}), one gets
\begin{eqnarray}
\label{c2} 
{\bbox {\Omega}} &=&\frac{m_1}{T_1}\pi^{-d} 
\sigma_{12}^{d-1}\mu_{21}(1+\alpha_{12})\beta_3n_2v_0^3\theta^{d/2}\int 
\,d{\bf V}_{1}^*\,\int \,d{\bf V}_{2}^*\,g^*{\bf g}_{12}^*
{\bf V}_{1}^*\nonumber\\
& & \times 
\left[1+{\sf C}_2:({\bf V}_{2}^*{\bf V}_{2}^*
-\case{1}{d}V_2^{*2}\openone)\right]
e^{-\theta V_{1}^{\ast 2}-V_{2}^{\ast 2}}\nonumber\\
&=&2n_{2}\pi^{-d}\mu_{21}(1+\alpha_{12})\sigma_{12}^{d-1}
\theta^{d/2+1}v_{0}\beta_3(1+\theta)^{-(1+d)}\int \,d{\bf x}\,\int 
\,d{\bf y}\,x\,{\bf x}({\bf x}+{\bf y}) \nonumber \\ 
& & \times \left[1+(1+\theta)^{-2}{\sf C}_2: 
({\bf y}-\theta {\bf x})({\bf y}-\theta {\bf x})\right]
e^{-bx^2-cy^2}, 
\end{eqnarray}
where $b=\theta(1+\theta)^{-1}$ and $c=(1+\theta)^{-1}$. This integral is 
easily performed, with the result
\begin{equation}
\label{c3} 
{\bbox {\Omega}}=
\frac{2}{d}\frac{\pi^{(d-1)/2}}{\Gamma(d/2)}\mu_{21}\sigma_{12}^{d-1}
n_2v_0(1+\alpha_{12})\left(\frac{1+\theta}{\theta}\right)^{1/2}
\left[\openone+\frac{1}{d+2}\frac{\theta}{1+\theta} \left(
{\sf P}_{2}^*-\openone\right)\right].
\end{equation} 
This expression yields Eq.\ (\ref{3.30}) given in the text.

\begin{figure}
\caption{Plot of the reduced elements of the pressure tensor of the 
granular gas as functions of the restitution coefficient $\alpha_{22}\equiv 
\alpha$ in the two-dimensional case. The solid lines correspond to the
present theory, the dotted ones to that of Ref.\ \protect\onlinecite{JR88}
and the dashed ones to that of Ref.\ \protect\onlinecite{BRM97}.
\label{fig1}}
\end{figure}

\begin{figure}
\caption{Dependence of the temperature ratio $\gamma\equiv T_1/T_2$ on the 
restitution coefficient $\alpha_{22}=\alpha_{12}\equiv \alpha$ in the 
two-dimensional case for $w=2$ and three different values of the mass ratio 
$\mu$. The dashed line corresponds to the prediction given by the theory of 
Jenkins and Mancini \protect\onlinecite{JM87} in the case $\mu=5$. 
\label{fig2}}
\end{figure}

\begin{figure}
\caption{Dependence of the diagonal and off-diagonal elements of the 
reduced self-diffusion tensor ${\sf D}^*$ on the restitution coefficient 
$\alpha_{22}=\alpha_{12}\equiv \alpha$ in the three-dimensional case.  
\label{fig3}}
\end{figure}

\begin{figure}
\caption{Dependence of the scalar (a) $\case{1}{3}D_{kk}^*$, the difference (b)
$D_{xx}^*-D_{yy}^*$, and the off-diagonal elements (c) $D_{xy}^*$ and (d)
$D_{yx}^*$ of the reduced tracer diffusion tensor ${\sf D}^*$ on the restitution coefficient 
$\alpha_{22}=\alpha_{12}\equiv \alpha$ in the three-dimensional case for 
$w=2$ and two values of the mass ratio $\mu$: $\mu=2$ (solid lines)
and $\mu=4$ (dashed lines).   
\label{fig4}}
\end{figure}

\begin{figure}
\caption{Plot of the scalar $\case{1}{3}D_{kk}^*$ and the reduced diffusion 
coefficient $D^*$ obtained in Ref.\ \protect\onlinecite{GD02}
as functions of the restitution coefficient
$\alpha_{22}=\alpha_{12}\equiv \alpha$ in the
three-dimensional case for $w=2$ and two values of the mass ratio $\mu$:
$\mu=2$ (solid lines) and $\mu=4$ (dashed lines).  
\label{fig5}}
\end{figure}

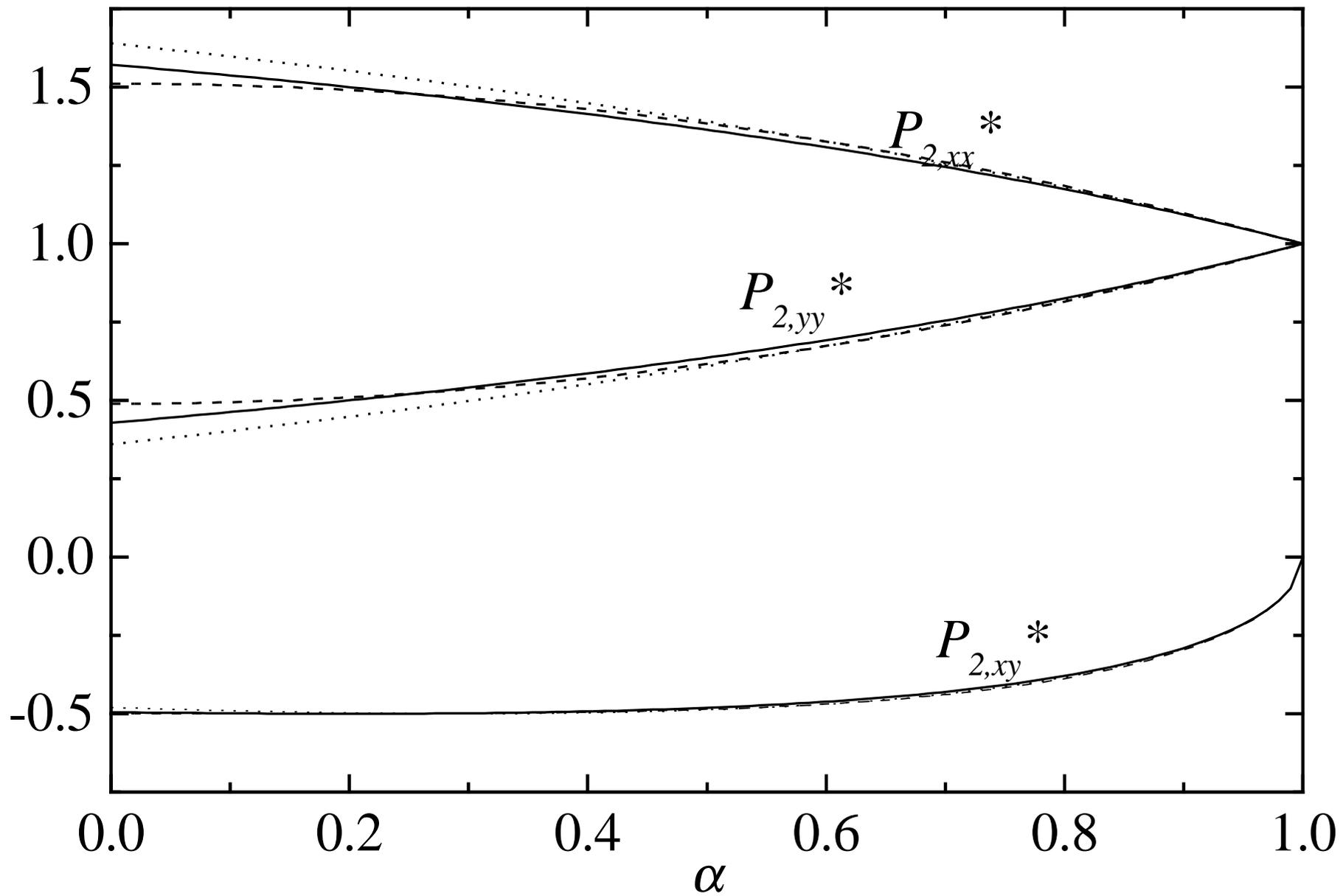
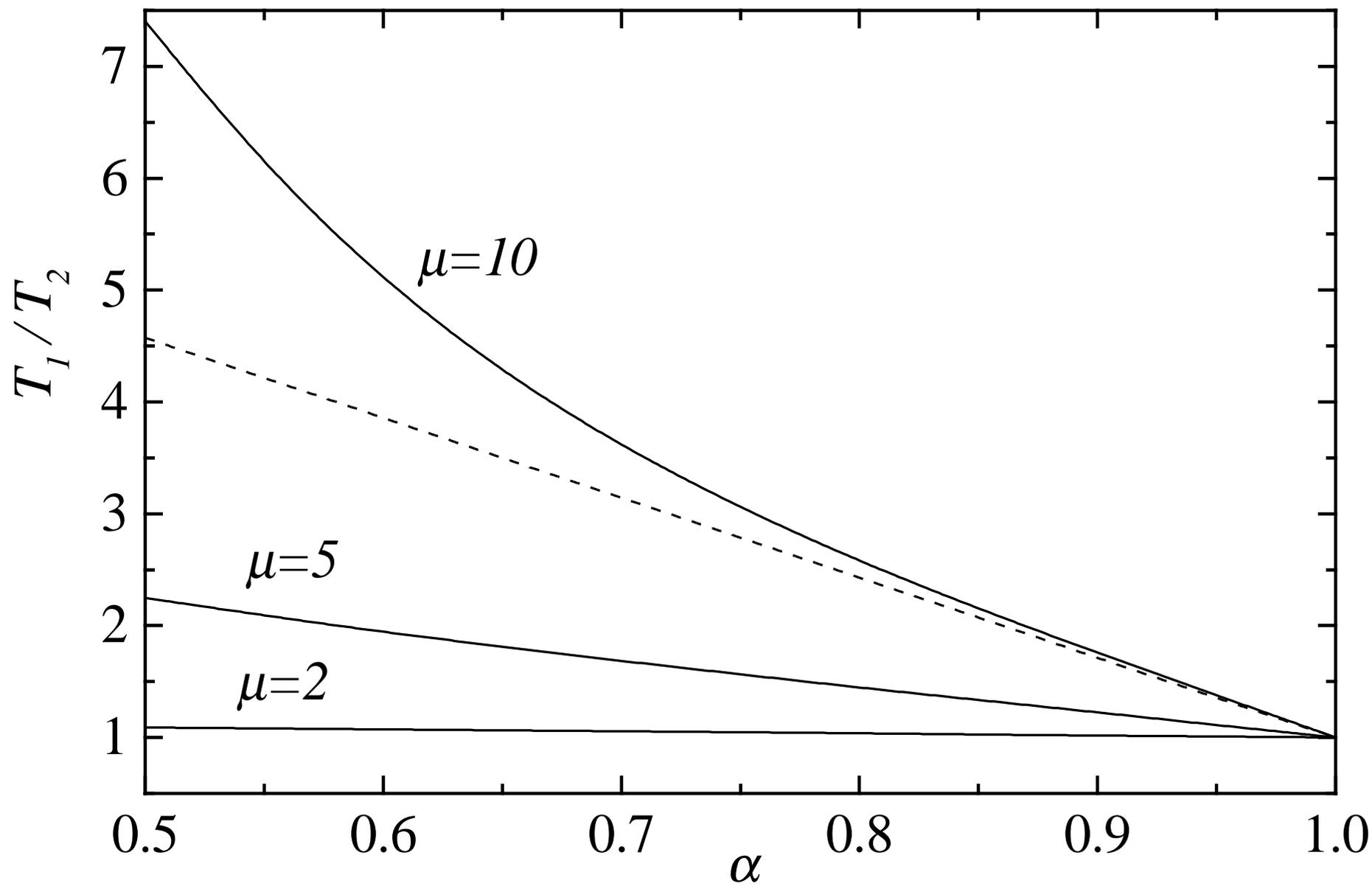
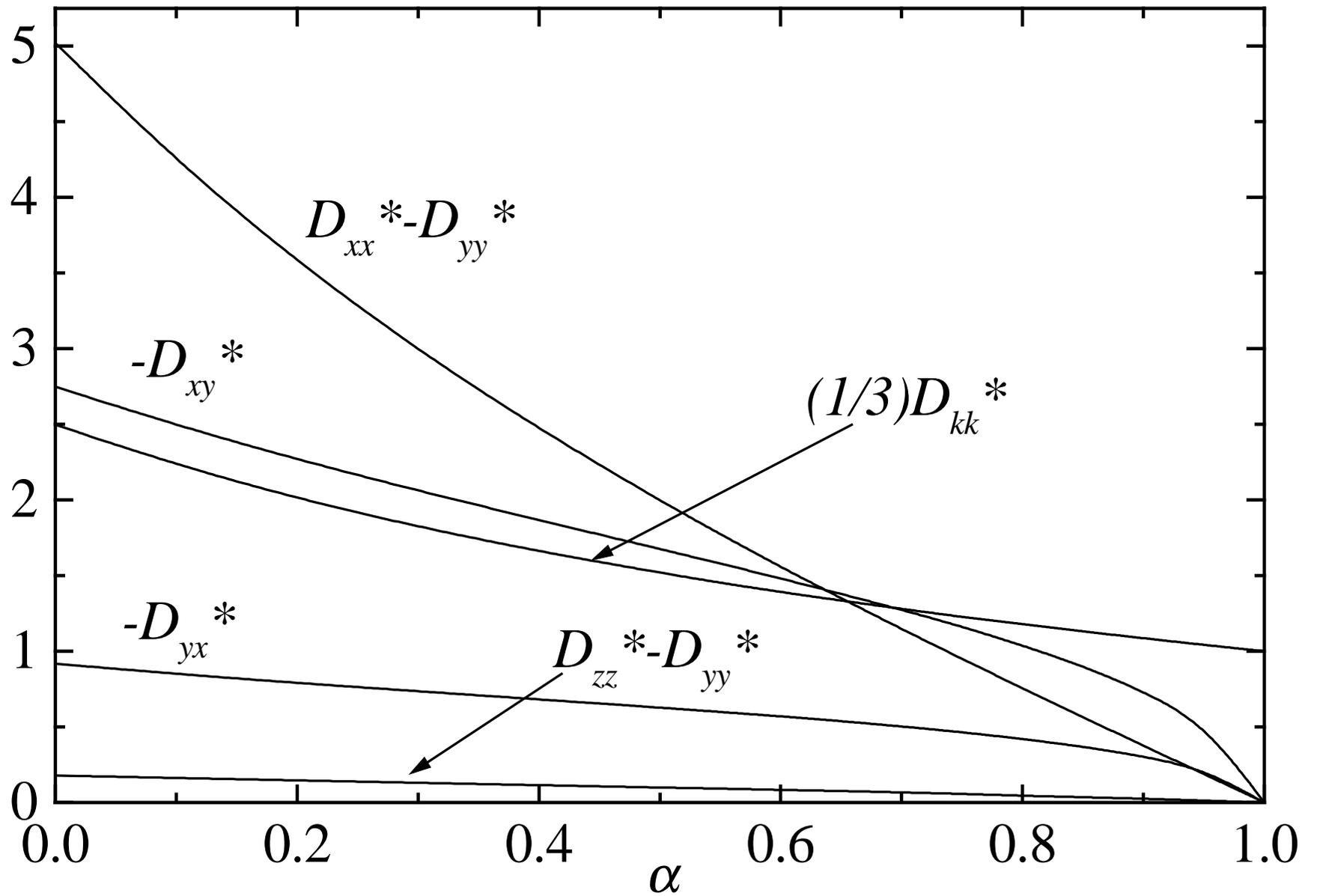
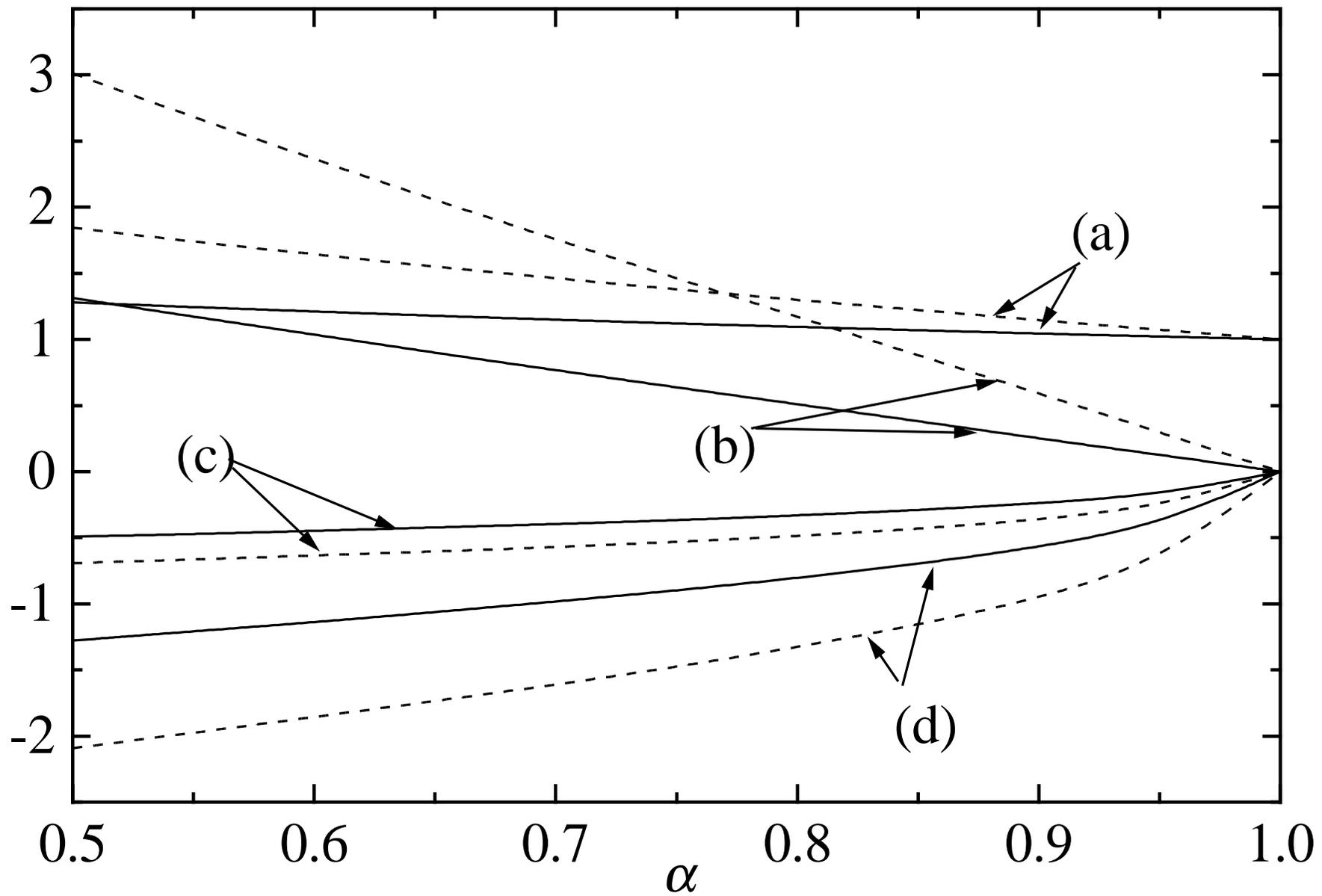
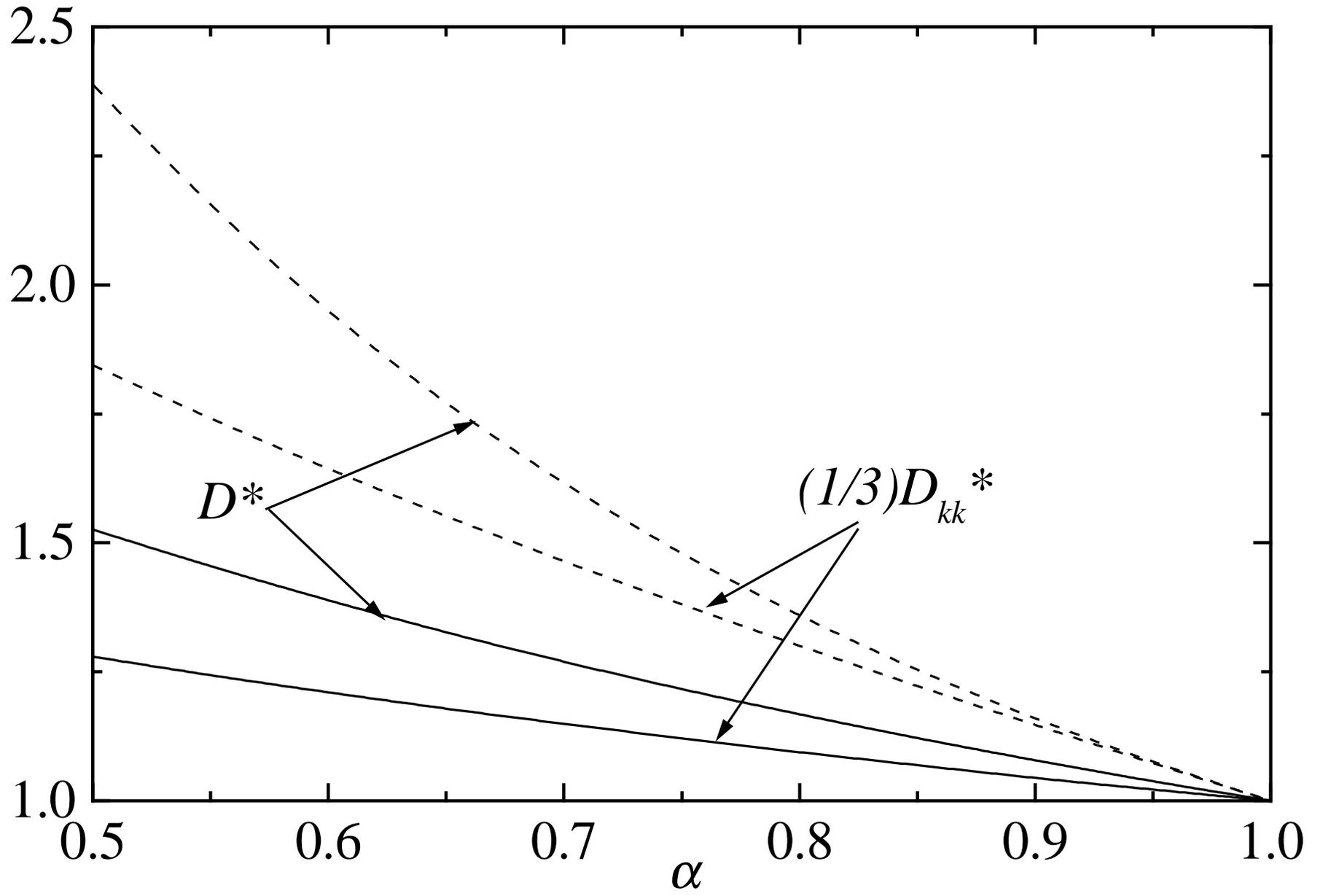
\end{document}